\documentclass[10pt,letterpaper,twocolumn]{article} 

\usepackage{ol2}
\usepackage[draft]{hyperref}
\usepackage{amsmath}

\begin{document}

\twocolumn[ 

\title{Non-Hermitian topological phase transitions in superlattices and the optical Dirac equation}


\author{Stefano Longhi}
\address{Dipartimento di Fisica, Politecnico di Milano and Istituto di Fotonica e Nanotecnologie del Consiglio Nazionale delle Ricerche, Piazza L. da Vinci 32, I-20133 Milano, Italy (stefano.longhi@polimi.it)}
\address{IFISC (UIB-CSIC), Instituto de Fisica Interdisciplinar y Sistemas Complejos, E-07122 Palma de Mallorca, Spain}

\begin{abstract}
Optical superlattices with sublattice symmetry subjected to a synthetic imaginary gauge field undergo a topological phase transition in the Bloch energy spectrum, characterized by the change of a spectral winding number. For a narrow gap, the phase transition is of universal form and described by a non-Hermitian Dirac equation with Lorentz-symmetry violation. A simple photonic system displaying such a phase transition is discussed, which is based on light coupling in co-propagating gratings.  
 
\end{abstract}

 ] 

{\em Introduction.}
Synthetic imaginary gauge fields \cite{r0} have found recently an increasing interest in photonics and beyond \cite{r1,r2,r3,r3b,r4,r5,r6,r7,r8,r9,r10,r11,referee,r12,r13,r15,r16,r17,r18,r19,r20,r21} owing to their ability to realize a non-Hermitian (NH) control of the flow of light \cite{r1,r2,r3,r12,r15,r17} and for the observation of nontrivial topological features in the energy band of NH systems, which underpin  notable phenomena such as the NH skin effect and a generalized bulk-edge correspondence \cite{r4,r5,r6,r7,r8,r9,r10,r11,referee,r12,r13,r15,r16,r17,r18,r19,r20,r21}. In a system with open boundary conditions (OBC), an imaginary gauge field $h$ does not change the energy spectrum from the Hermitian limit $h=0$, while all bulk modes are squeezed towards the edge (see e.g. \cite{referee}). On the other hand, under periodic boundary conditions (PBC) the energy spectrum becomes complex and exhibits nontrivial topological features characterized by integer nonzero winding numbers. This mens that, while under OBC the NH Hamiltonian is topologically equivalent to the Hermitian one, this is not the case of a system with PBC. An important example of system displaying non-trivial topology is a binary lattice with sublattice (chiral) symmetry, the prototypal model being the famous  Su-Schrieffer-Heeger (SSH) model of polyacetylene \cite{r22} and its extension including long-range hopping respecting sublattice symmetry \cite{r23}. In the Hermitian limit, in the nontrivial topological phase such model exhibits zero-energy edge states, whose number is provided by a topological number $\mathcal{W}$ according to the bulk-boundary correspondence \cite{r23}. Several NH extensions of the SSH model have been investigated in recent works \cite{r3,r6,r8,r24,r25,r26,r27,r28}, including the case of asymmetric hopping corresponding to the application of an imaginary gauge field \cite{r3,r6}. Asymmetric hoppings have been realized in different photonic settings, such as in microring chains with lossy auxiliary rings \cite{r12} and in synthetic mesh lattices \cite{r15}.\\
In this Letter we unravel topological phase transitions in the Bloch band energy spectrum of binary superlattices with sublattice symmetry (SLS) under an imaginary gauge field, characterized by the change of a spectral winding number $\mathcal{W}_s$. For a narrow gap, the phase transition is of universal form and described by a NH Dirac equation with Lorentz-symmetry violation, which is obtained from the tight-binding model in the long-wavelength limit \cite{r29,r30,r31}.  A simple photonic system displaying such a phase transition in the continuous limit is finally discussed, which is based on grating-assisted codirectional coupling of light in two waveguides with loss and/or gain.

{\em Model and non-Hermitian topological phase transition.} We consider a binary lattice with SLS and possible long-range hopping \cite{r23} with an applied imaginary gauge field $h$ \cite{r0}. In physical space, the system is described by tight-binding equations for the amplitudes $a_n(t)$ and $b_n(t)$ in the two sublattices A and B
\begin{eqnarray}
i \frac{d a_n}{dt} &= & \sum_{l=-N+1}^N \rho_l b_{n-l} \exp(-h l) \\
i \frac{d b_n}{dt} &= & \sum_{l=-N+1}^N \rho_l^* a_{n+l} \exp(h l)
\end{eqnarray}
where $\rho_l$ is the (Hermitian) hopping amplitude between sites $a_n $ and $b_{n-l}$, and $N$ is the maximum non-negligible long-range hopping. The usual SSH model ($N=1$) is attained by letting $\rho_0=t_1$ and $\rho_1=t_2$ [see inset in Fig.1(a)]. 
 In a system with OBC, the imaginary gauge field can be eliminated by the non-unitary gauge transformation  \cite{r3,referee} $a_n= a^{\prime}_n \exp(-nh)$, $b_n= b^{\prime}_n \exp(-nh)$, and thus the energy spectrum is not modified by the gauge field while all bulk eigenstates are squeezed toward the edge (skin effect). This means that under OBC the NH Hamiltonian in physical space is topologically equivalent to the one in the Hermitian limit $h=0$. Here we focus our attention to the PBC case, where the imaginary gauge field induces a topological phase transition as discussed below. 
  The Bloch Hamiltonian is described by the $2 \times 2$ matrix $H(k)$
  with elements $H_{11}=H_{22}=0$, $H_{12}=P(z)$, $H_{21}=Q(1/z)$, where we have set
  $z=\exp(-ik-h)$ and
\[ P(z)=\sum_{l=-N+1}^{N} \rho_l z^l \; , \; Q(z)= \sum_{l=-N+1}^{N} \rho_l^* z^l .\]
The system exhibits SLS, i.e. $\sigma_z H(k)=-H(k) \sigma_z$ where $\sigma_z$ is the Pauli matrix. This means that the energy spectrum under PBC is symmetric around the zero-energy point $E_F=0$ and reads
\begin{equation}
E_{\pm}(k)= \pm \sqrt{P(z) Q(1/z)}.
\end{equation}
The central concept here is that of topological equivalence of Hamiltonians and topological phase transitions. Let $H(\lambda)$ be a family of Hamiltonians with SLS depending on a parameter $\lambda$, and let us assume that at $\lambda=\lambda_{1,2}$ the zero-energy $E_F=0$ does not belong to the energy spectrum of neither $H_1=H(\lambda_1)$ and $H_2=H(\lambda_2)$, i.e. $E_F$ is a point-gap of both $H_1$ and $H_2$. The two Hamiltonians are topologically equivalent if and only if by varying $\lambda$ they can be continuously deformed into each other while retaining SLS and the point-gap energy $E_F=0$ \cite{r11}. The Hermitian limit ($h=0$) is well know: a topological phase transition is signaled by the change of the winding number $\mathcal{W}$, that describes the times the vector $P(z)$ encircles the origin in complex plane as the Bloch wave number $k$ spans the first Brillouin zone, from $k=-\pi$ to $k=\pi$. The bulk-boundary correspondence ensures that, for a non vanishing $\mathcal{W}$, there exist exactly $| \mathcal{W} |$ pairs of zero-energy edge states, the largest number of $ | \mathcal{W} |$ being $N$ \cite{r23}. For example, in the usual SSH model, for a given value of $t_1$ and assuming the hopping amplitude $\lambda=t_2$ as the varying parameter, a topological phase transition occurs at $t_2=t_1$, where the gap closes. Here we focus our attention to the NH case, where the hopping amplitudes in the model are fixed while the family (varying) parameter is the imaginary gauge field, i.e. $\lambda=h$. Clearly, under OBC the energy spectrum does not depend on $h$, and the non-unitary gauge transformation mentioned above ensures that the number of pairs of zero-energy edge states is provided again by the winding $| \mathcal{W}|$, while all bulk states are squeezed to the left edge (skin effect). Accidentally, at special values of $h$ some zero-energy edge states could become delocalized, as discussed in \cite{r3}. Conversely, under PBC the energy spectrum depends  on $h$ and its topology is described by a spectral winding number $\mathcal{W}_s$ with respect to a base energy $E_F$ \cite{r3,r13,r18}. For a gapped system with SLS, it is worth considering $E_F=0$ (the Fermi energy in the Hermitian limit), so as $\mathcal{W}_s$ reads
\begin{equation}
\mathcal{W}_s= \frac{1}{2 \pi i} \int_{-\pi}^{\pi}
 dk \frac{d}{d k} \log \det H(k) = \frac{1}{2 \pi i} \sum_{l= \pm} \int_{-\pi}^{\pi}.
 dk \frac{d}{d k} \log E_l(k)
 \end{equation}
  Physically, $ |\mathcal{W}_s|$ corresponds to the number of edge states of the lattice with energy $E_F=0$ under semi-infinite boundary conditions \cite{r13}.
 \begin{figure}[htb]
\centerline{\includegraphics[width=8.7cm]{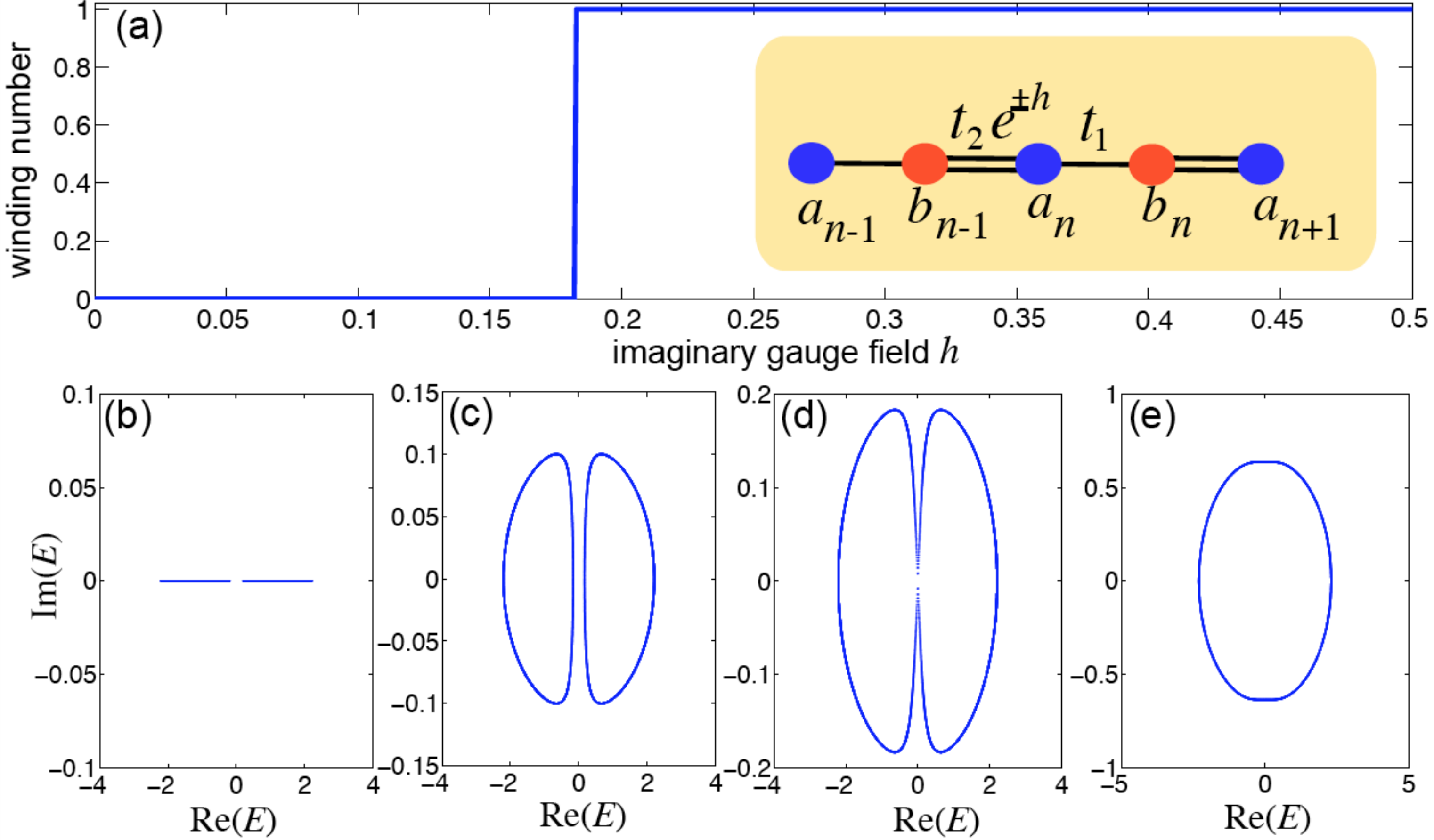}} \caption{ \small
(Color online) Topological spectral phase transition in the ordinary SSH for parameter values $\rho_0=t_1=1$ and $\rho_1=t_2=1.2$. (a) Behavior of the winding number $\mathcal{W}_s$ versus imaginary gauge field $h$. Note the phase transition at $h=h_1= | \log (t_2 / t_1)| \ \simeq 0.1823$. The inset in (a) shows a schematic of the SSH model with the imaginary gauge field. (b-e). PBC energy spectrum in complex plane of $H(k)$ for a few increasing values of $h$: (b) $h=0$, (c) $h=0.1$, (d) $h=h_1$, and (e) $h=0.6$.}
\end{figure} 
\begin{figure}[htb]
\centerline{\includegraphics[width=8.7cm]{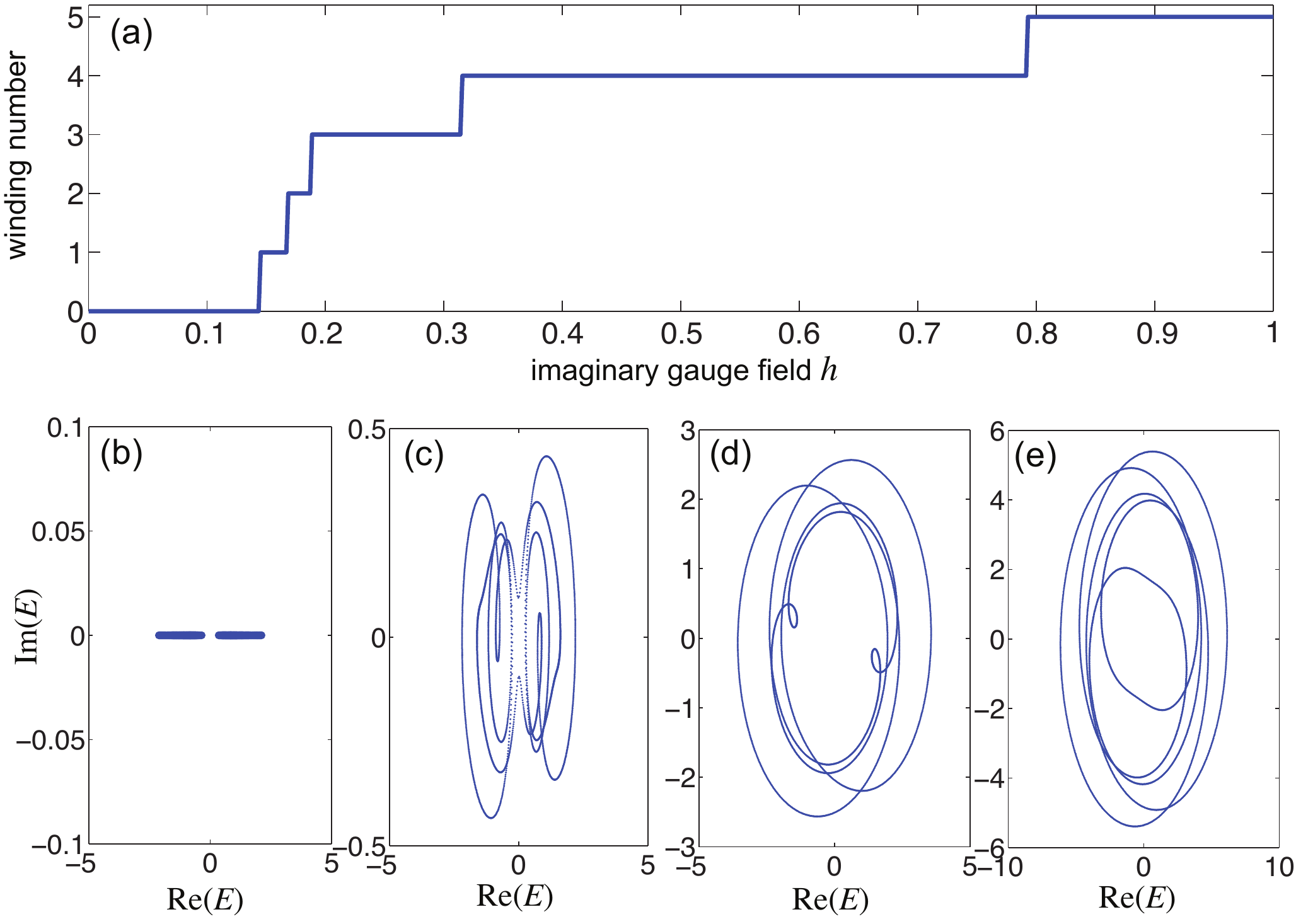}} \caption{ \small
(Color online) Same as Fig.1, but for the SSH with long-range hopping ($N=3$). Parameter values are $\rho_{-2}=0.2$, $\rho_{-1}=0.3$, $\rho_0=0.2i$, $\rho_1=0.5$, $\rho_2=0.1$ and $\rho_3=1$. In this case, as $h$ is increased, there are $(2N-1)=5$ discontinuities of the spectral winding at the values $h_1=0.1449$, $h_2=0.1684$, $h_3=0.1885$, $h_4=0.3149$ and $h_5=0.7928$ [according to Eq.(5)]. The PBC energy spectra in (b-e) are computed for (b) $h=0$, (c) $h=0.15$, (d) $h=0.6$, and (e) $h=0.9$.}
\end{figure}
At $h=0$ (Hermitian limit) we assume that the system is gapped, i.e. $E_F=0$ does not belong to the energy spectrum of $H$; clearly, $W_s=0$ since the spectrum is entirely real. As the imaginary gauge field $h$ is increased, a sequence of spectral topological phase transitions is observed, at which the point-gap $E_F=0$ closes and $\mathcal{W}_s$ changes by one unity. It can be readily shown that the phase transitions occur at the values of $h$ given by 
\begin{equation}
h_l=\left| \ \log | z_l| \right|
\end{equation}
 ($l=1,2,...,2N-1$), where $z_l$ are the $(2N-1)$ roots  of $P(z)$. In fact, at such values of $h$ there exists a Bloch wave number $k=\tilde{k}$ such that $P(z)$ (or $Q(1/z)$) vanishes, corresponding to $E_{\pm}(\tilde{k})=0$ and undefined winding $W_s$, i.e. a discontinuity of $W_s$ versus $h$.  Moreover, at such values of $h$, $H(\tilde{k})$ is defective and thus $E_F=0$ is an exceptional point.
 The largest value of $W_s$ is given by $(2N-1)$. In fact, for large $h$ one has asymptotically $E_{\pm} \sim  \pm \sqrt{\rho_{-N+1}\rho^*_N} \exp [i (2N-1) (k-ih)/2]$, corresponding to a winding $W_s=(2N-1)$.  For example, in the standard SSH model $P(z)=t_1+t_2 z$, corresponding to a single root $z_1=-t_1/t_2$ and a phase transition point $h=h_1=| \log (t_2/t_1)|$ (see Fig.1). In the presence of long-range hopping, a larger number of spectral phase transitions is found. As an illustrative example, Fig.2 shows the cascade of phase transitions for $N=3$. We stress that such spectral topological phase transitions do not correspond to the appearance or disappearance of zero-energy edge states, which remain unchanged as $h$ is varied and are determined by the value of the Hermitian winding number $\mathcal{W}$ solely.
 
 {\em Normal form of the phase transition and the non-Hermitian Dirac model.}
Let us assume that in the Hermitian limit $h=0$ the energy spectrum of $H(k)$ shows a narrow gap at the Bloch wave number $k=k_0$ of width $2 \Delta$. This means that a root to $P(z)$, says $z=z_1$, has a modulus close to one.  At around $k=k_0$, the dispersion curves Eq.(3)
can be thus approximated by the hyperbolic curves describing an avoided crossing
\begin{equation}
E_{\pm}(k) \sim \pm \sqrt{\Delta^2+ \beta (k-k_0)^2}
\end{equation}
with $\beta>0$ related to the curvature (effective mass) of the dispersion curves at $k=k_0$. When a small imaginary gauge field is applied, the dispersion relations are simply obtained from Eq.(6) after the replacement $k \rightarrow k-ih$, which thus provides the normal form of the energy spectrum near the phase transition point in a narrow-gap system. The spectral topological phase transition, as $h$ is slightly increased above zero, corresponds to the touching of the two dispersion curves in the complex energy plane at the point gap $E_F=0$, which occurs at $k=k_0$ for 
\begin{equation}
h=h_1=\Delta / \sqrt{\beta}. 
\end{equation}
Such a general result is illustrated in Fig.3 for the standard SSH model, displaying a small gap ($t_1 \sim t_2$) at $k_0= \pi$. In this case, $\Delta=|t_2-t_1|$, $\beta=t_1 t_2$, and the critical value of the gauge field $h_1$ at the phase transition, obtained from Eq.(7), is equivalent to exact result [Eq.(5)] in the $t_2 \simeq t_1$ limit. 
Interestingly, the normal form of the phase transition in the small-gap limit, governed by the hyperbolic form (6) of the dispersion relation, can be traced back to a NH extension of the Dirac equation with Lorentz-symmetry violation \cite{r32,r33}, which is obtained from the tight-binding model (1) in the long-wavelength approximation \cite{r29,r30,r31}.  To this aim, we make in Eq.(1) the Ansatz $(a_n,b_n)^T=(A(n),B(n))^T \exp(i k_0 n)$ with $A(n), B(n)$ slowly-varying envelopes with respect to $n$. For a small gauge field $h$, after Taylor expansion from Eq.(1) it readily follows that the envelopes $A(n), B(n)$ satisfy coupled-mode equations that can be cast in the form of a Dirac equation \cite{r29} with a NH term. To write the Dirac equation in the Weyl (or chiral) form, it is worth introducting the variable transformation  $\psi_1=1/ \sqrt{2} [ B \exp (i \varphi /2)-i A \exp(-i \varphi /2)]$, $\psi_2=1/ \sqrt{2} [ A \exp (-i \varphi /2)-i B \exp(i \varphi /2)]$ so that $\psi=(\psi_1, \psi_2)^T$ satisfies  the Dirac-Weyl equation 
\begin{equation}
i \partial_t \psi=  \sigma_z R_2 ( i \partial_n+i h) \psi+ R_1 \sigma_x \psi,
\end{equation}
where $\sigma_{x,z}$ are the Pauli matrices, and the real parameters $R_1$, $R_2$ and $\varphi$ are defined by the relations $\sum_{l} \rho_l \exp(-i k_0l) \equiv R_1 \exp(i \varphi)$ and 
$\sum_{l} l \rho_l \exp(-i k_0l) \equiv R_2 \exp(i \varphi)$. The amplitudes $R_1$ and $R_2$ are related to the gap size $2 \Delta$ and band curvature $\beta$ at $k=k_0$ by the relations $\beta=R_2^2$ and $\Delta=R_1$.
\begin{figure}[htb]
\centerline{\includegraphics[width=8.7cm]{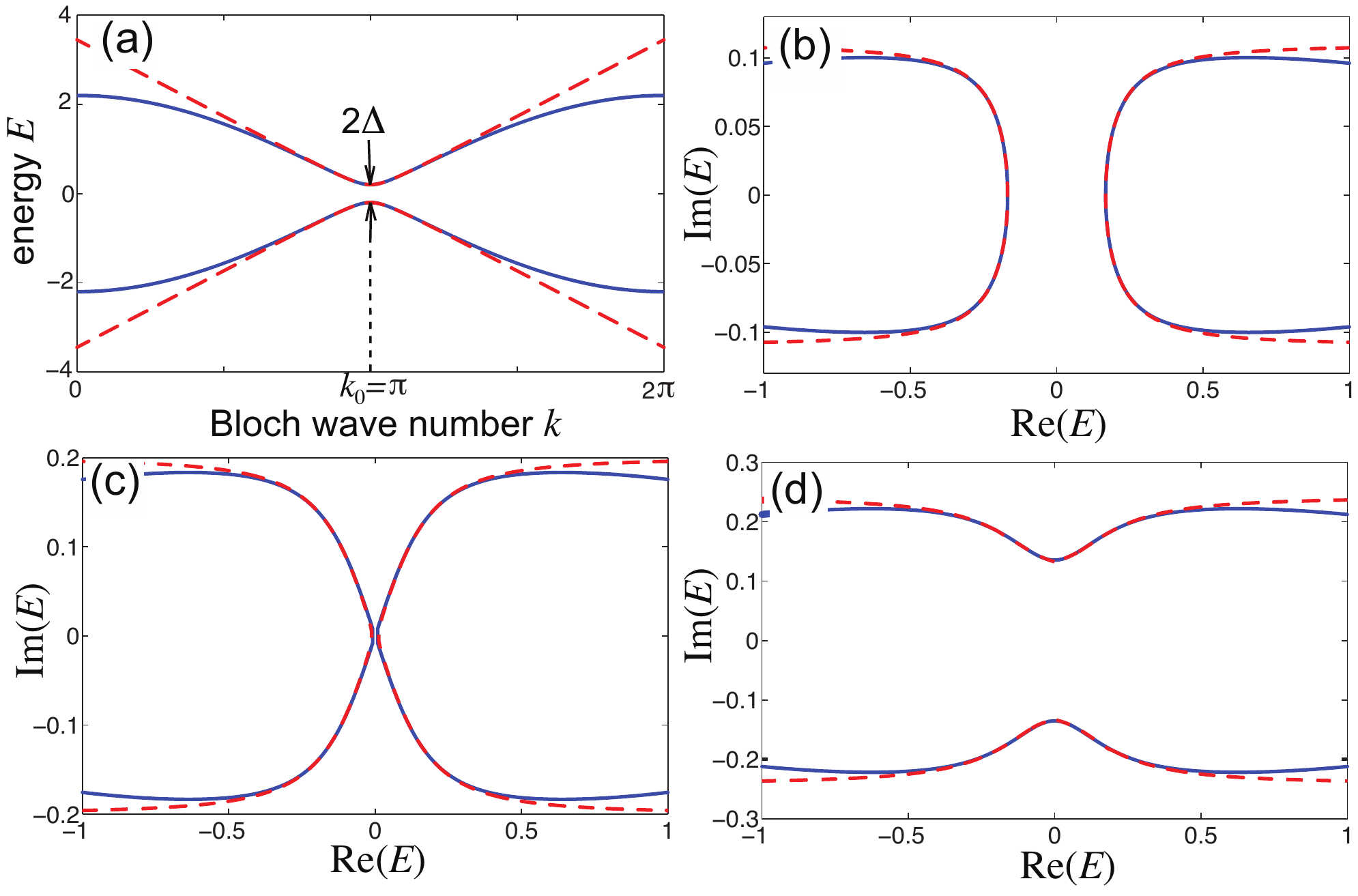}} \caption{ \small
(Color online) Normal form of the topological phase transition in the small-gap limit, illustrated for the standard SSH model. (a) Dispersion relation of the SSH model in the Hermitian limit ($h=0$) for $t_1=1$ and $t_2=1.2$. A small gap of width $2 \Delta=2 |t_2-t_1|$ occurs at the wave number $k_0= \pi$. The dashed curves show the hyperbolic approximation  of the dispersion curves near the gap [Eq.(6)]. (b-d) Energy spectrum in complex plane for a few increasing values of $h$: (b) $h=0.1$, (c) $h=h_1=0.1823$ (phase transition point), and (d) $h=0.22$.}
\end{figure}
Note that the Dirac Hamiltonian in Eq.(8) is NH for $h \neq 0$, and such NH term breaks the Lorentz symmetry \cite{r32,r33}. The energy spectrum of the NH Dirac Hamiltonian is complex and formed by two energy branches, which have the normal form given by Eq.(6). Hence the the spectral topological phase transition of a narrow-gap superlattice with SLS can be described rather generally by a NH Dirac equation with Lorentz symmetry violation.\par
  
{\em Light coupling in co-propagating gratings with gain and loss.}
Light propagation in periodic (Bragg) structures or in nonlinear second-order optical media provide an experimentally accessible platform to emulate in photonics relativistic wave equations \cite{r29,r30,r32}, including NH relativistic models \cite{r34}. To realize the NH Dirac-Weyl model Eq.(8) with Lorentz-symmetry violation, we consider a long-period grating (LPG) structure (see e.g. \cite{r35,r36,r37,r38}), in which
selective energy transfer between two nonidentical (i.e., non-synchronous) co-propagating waveguide
modes is obtained via a periodic coupling along the propagation direction $z$, shown schematically in Fig.4(a). The grating period is chosen to
satisfy the phase-matching condition at the carrier (reference) frequency $\omega_0$. Indicating by $v_1$ and $v_2$ the group velocities of the two waveguide modes at the carrier frequency and by $1/v=(1/2)(1/v_1+1/v_2)$ the mean of the inverse of group velocities, in the reference frame $(z,\tau)$ with retarded time $\tau=t-z/v$ the coupled-mode equations for the mode envelopes $\psi_1(z,\tau)$ and $\psi_2(z,\tau)$ read \cite{r31}
\begin{eqnarray}
i \left( \partial_z \psi_1-\delta \partial_ \tau \psi_1 \right)& = & -i \gamma_1 \psi_1+ \kappa \psi_2 \\
i \left( \partial_z \psi_2+\delta \partial_ \tau \psi_2 \right) & = & -i \gamma_2 \psi_2+\kappa \psi_1
\end{eqnarray}
where $\kappa$ is the coupling constant, $\delta=(1/2)(1/v_2-1/v_1)$ accounts for the group velocity mismatch of the two modes, and $\gamma_{1,2}$ are the propagation losses (or gain for $\gamma_{1,2}<0$) in the two waveguides. 
Clearly, for a balanced gain-loss system, with $\gamma_2=-\gamma_1 \equiv \gamma_0 >0$,  Eqs.(11-12) reproduce the NH Dirac model (8) provided that the following substitutions are made:  $z \rightarrow t$, $ \tau \rightarrow n$, $\delta \rightarrow R_2=\sqrt{\beta}$, $\kappa \rightarrow R_1$ and $\gamma_0 \rightarrow R_2 h$. Hence the modal gain/loss term $\gamma_0$ in the LPG is responsible for the NH term in the Dirac equation that breaks Lorentz invariance, which is distinct than other NH Dirac models with an imaginary mass term induced by a combined gain/loss grating \cite{r34}, where Lorentz symmetry is not broken. For a purely dissipative codirectional coupler ($\gamma_{1,2} \geq 0$), the equivalence with the NH Dirac model is still valid provided that the gauge transformation $\psi_{1,2} \rightarrow \psi_{1,2} \exp [-(\gamma_1+\gamma_2)z/2]$ is performed.\\ A main question is whether a signature of the spectral phase transition, described by Eq.(8), can be detected from simple transmission experiments. 
An important parameter of a LPG device, when used as a band-rejection filter \cite{r37,r38}, is the spectral transmission $t(\omega)=\psi_1(L) / \psi_1(0)$, when $L$ is the interaction length the input field in waveguide 1 is monochromatic with frequency $\omega$ close to $\omega_0$ \cite{r35,r36,r37,r38}.  From the coupled-mode equations the expression of $t(\omega)$ can be readily obtained as
\begin{equation}
t(\omega)= \left\{  \cos(E L)-i \frac{\sigma}{E} \sin (EL) \right\} \exp [-(\gamma_1+\gamma_2)L/2]
\end{equation} 
\begin{figure}[htb]
\centerline{\includegraphics[width=8.7cm]{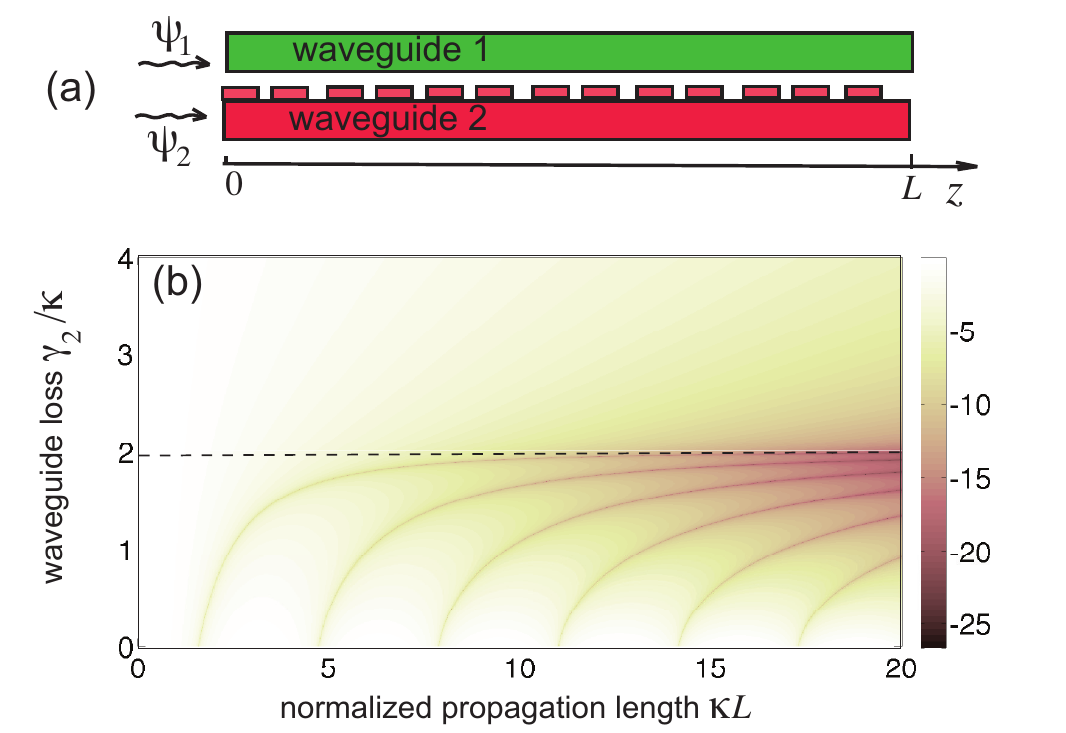}} \caption{ \small
(Color online) (a) Schematic of grating-assisted codirectional coupling between the optical modes in two waveguides 1 and 2. The loss (gain) coefficient in the two guides is $\gamma_1$ and $\gamma_2$, while the effective mode coupling is $\kappa$. In a codirectional coupler with balanced gain and loss, $\gamma_2=-\gamma_1>0$,whereas in a purely passive coupler  $\gamma_{1,2} \geq 0$. (b) Behavior of the transmission amplitude $|t(\omega_0)|$ of the codirectional coupler at resonance $\omega=\omega_0$, depicted on a log scale in a pseudo color map, as a function of waveguide loss $\gamma_2$ (with $\gamma_1=0$) and propagation length $L$. The dashed horizontal line corresponds to the phase transition $\kappa=\gamma_2 /2$ in the NH Dirac model. The dark curves below the dashed line correspond to the vanishing of the transmission amplitude.}
\end{figure}
where we have set $E= \sqrt{\kappa^2+ \sigma^2}$, $\sigma= i \gamma- (\omega-\omega_0) \delta$, and $\gamma=(\gamma_2 -\gamma_1)/2$. 
 The condition of perfect signal rejection from the LPG filter at resonance corresponds to $t(\omega_0)=0$, and has been discussed in previous works \cite{r37}. In the presence of losses, such condition reads 
\begin{equation}
\gamma \tan \left( \sqrt{\kappa^2-\gamma^2} \; L \right)= - { \sqrt{\kappa^2-\gamma^2}}.
\end{equation}
For given values of the coupling $\kappa$ and loss unbalance $\gamma$>0, Eq.(12) can be satisfied for an uncountable set of lengths $L$ provided that $\kappa > \gamma$, while it cannot be satisfied when $\kappa < \gamma$. Hence, the topological phase transition of the NH Dirac model is signaled by a qualitative change in the transmittance of the LPG structure, as shown in Fig.4(b): In the topological phase $\mathcal{W}_s=0$ ($\kappa > \gamma$), perfect signal rejection is possible, indicated by the black curves in Fig.4(b), while in the topological phase $\mathcal{W}_s=1$ ($\kappa < \gamma$) perfect signal rejection is prevented. 

{\em Conclusions.} In this work we unravelled a general route to non-Hermitian topological phase transitions in the Bloch energy spectrum of superlattices with sublattice symmetry under a synthetic imaginary gauge field. In the narrow-gap limit, the phase transition shows a universal form, described by a NH Dirac-Weyl equation with Lorentz-symmetry violation obtained from the tight-binding model in the long-wavelength  limit. Since the NH Dirac-Weyl equation can describe several continuous optical models, from grating-assisted co-directional coupling of light discussed in this work to nonlinear processes as sum-frequency generation \cite{r31}, our results advance the frontiers of NH topology in optical systems, beyond the usual tight-binding models, and suggest a simple route to implement synthetic imaginary gauge fields in the long-wavelength (continuous) approximations of lattice models.\\

\end{document}